# Analyzing Onset of Nonlinearity of a Colloidal Gel at the Critical Point

Khushboo Suman and Yogesh M. Joshi*

Department of Chemical Engineering,

Indian Institute of Technology Kanpur 208016, India

* Corresponding Author, E-mail: joshi@iitk.ac.in,

Tel.: +91 512 2597993, Fax: +91 512 2590104

**Abstract**

In this work we study onset of nonlinear rheological behavior of a colloidal dispersion of a synthetic hectorite clay, LAPONITE®, at the critical gel state while undergoing sol-gel transition. When subjected to step strain in the nonlinear regime, the relaxation modulus shifts vertically to the lower values such that the deviation from linearity can be accommodated using a strain dependent damping function. We also perform creep-recovery and start-up shear experiments on the studied colloidal dispersion at the critical gel state and monitor deviation in response as the flow becomes nonlinear. A quasi-linear integral model is developed with the time – strain separable relaxation modulus to account for the effect of nonlinear deformation. Remarkably, the proposed model predicts the deviation from linearity in the creep-recovery and start-up shear experiments very well leading to a simple formulation to analyze the onset of nonlinear rheological behavior in the critical gels. We also analyze the energy dissipation during the nonlinear deformation and validate the Bailey criterion using the developed viscoelastic framework.

# I. Introduction:

A hierarchal fractal structure that is devoid of any single prominent length-scale as well as time-scale is often observed in various kinds of soft materials [1]. The typical examples of such structure are various kinds of chemical [2,3] and physical gels [4-7], protein gels [8,9], carbohydrates [10,11], long-chain branched polymer [12] and human tissues [13,14] at the critical point. Interestingly, the critical gel behavior is exhibited by many food ingredients like wheat flour dough [15,16], egg yolks [17], milk casein [18,19] and carrageenan solution [20,21]. Such scale-free topological structure is defined by the largest aggregate or a polymeric network that spans the entire space resulting in a three-dimensional weakest percolated network [22]. Rheologically such structure gets manifested in power-law dependence of the response functions such as elastic and viscous moduli, relaxation modulus and compliance on their respective independent variables, frequency or time in a limit of linear viscoelasticity [23]. As the applied deformation field gets into the nonlinear domain, the network may get stretched and eventually ruptured. A huge amount of work has been carried out in the literature over the past three decades on linear viscoelastic behavior of the critical gels composed of the crosslinked polymer network [3,12] and nonlinear rheological behavior of the same [15,19,24-26]. The colloidal gels that are formed with much weaker interaction energy of the interparticle bonds (van-der Waals, electrostatic, depletion, hydrogen bonding, etc.) compared to polymeric (chemical) gels (covalent bonds), have been much less explored in the literature. In this work, we study linear and nonlinear viscoelastic behavior of aqueous synthetic hectorite clay dispersion at the critical gel state. We also investigate how the response to nonlinear deformation field can be predicted from the linear viscoelastic constitutive framework.

For the polymeric materials undergoing a crosslinking reaction, in a seminal contribution, Winter and Chambon [3,27] described the viscoelastic properties of a unique state where the weakest space spanning network gets formed. They observed that at the critical gel state, which is independent of observation timescale, the elastic $(G')$ and viscous $(G'')$ moduli exhibit identical dependence on angular frequency $(\omega)$ over a wide range of frequencies as given by [3]:

$$G' = G'' \cot\left(n_c \pi/2\right) = \frac{\pi S}{2\Gamma(n_c)\sin\left(n_c\pi/2\right)}\omega^{n_c}, \tag{1}$$

where $\Gamma(n_c)$ is the Euler gamma function of the critical relaxation exponent $n_c$ having limits $0 \leq n_c \leq 1$ and $S$ is the gel network strength. As a result, the damping factor, given by $\tan\delta = \tan\left(n_c\pi/2\right) = G''/G'$, becomes independent of the applied angular frequency. Furthermore, at the critical gel state, upon the application of step strain in a linear response limit, the stress relaxation modulus $G(t)$ is observed to show a power-law dependence on time as given by [24]:

$$G(t) = St^{-n_c}. \tag{2}$$

Equivalently, application of creep (constant stress) in a limit of linear viscoelasticity, leads to compliance given by [24]:

$$J(t) = \frac{(1-n_c)}{S\Gamma(2-n_c)\Gamma(1+n_c)}t^{n_c}. \tag{3}$$

The material at the critical gel state can be characterised by the fundamental law of linear viscoelasticity: the Boltzmann superposition principle (BSP), given by [28]:

$$\underset{\approx}{\sigma}(t) = \int_0^t G(t-t')\underset{\approx}{\dot{\gamma}}(t')dt', \tag{4}$$

where $\underset{\approx}{\sigma}$ is the stress tensor, $\underset{\approx}{\dot{\gamma}}$ is the rate of strain tensor, $t'$ is the time of application of the deformation field, and $t$ is the present time. Incorporating the specific form of relaxation modulus at the critical gel state given by Eq. (2) into Eq. (4) leads to the constitutive equation given by [24]:

$$\underset{\approx}{\sigma}(t) = S\int_0^t (t-t')^{-n_c}\underset{\approx}{\dot{\gamma}}(t')dt'. \tag{5}$$

Furthermore, the use of convolution equation $t = \int_0^t G(s)J(t-s)ds$ and Eq. (2) leads to expression of compliance given by Eq. (3), while subjecting constitutive equation to $\gamma = \gamma_0 \sin \omega t$ results in $G'(\omega)$ and $G''(\omega)$ given by Eq. (1). These criterions proposed by Winter have been observed to be followed by polymeric materials [3,29-33], colloidal dispersions [7,34] and foodstuff [10,16,35] undergoing the sol-gel transition.

The power-law rheology gives rise to a power-law distribution of the relaxation time spectrum with a negative exponent $n_c$ given by: $H(\tau) \sim \tau^{-n_c}$. Such dependence arises from the hierarchical fractal behavior of the critical gel state, which is devoid of any characteristic time-scale or length-scale [23]. This suggests that the 'self-similar' behavior of the critical gel leads to the identical behavior in relaxation of stress at all scales of observation thus leading to 'scale invariance' [23]. Such dependence indicates the natural simplicity and elegance of the material at the critical gel state that is suggestive of a fractal structure. For the power-law rheology, observed at the critical state, Muthukumar [36] related $n_c$ to the fractal dimension $(f_d)$ as $f_d = 5(2n_c - 3)/2(n_c - 3)$. This relationship has been experimentally validated for fractal systems by conducting simultaneous scattering and rheological measurements [6,37,38].

With an increase in strain, nonlinearities such as stretching of a network and/or eventual breakage of the same cause alteration of the response functions from the expressions given by Eqs. (1) to (3). Consequently, Eq. (5) cannot be used to represent the critical gel state undergoing a nonlinear deformation. The study of the nonlinear deformation of a critical gel plays an important role in wide-ranging applications from kneading of dough to failure strain of an adhesive. However, modelling the material in the nonlinear regime poses a considerable challenge as it becomes necessary to use nonlinear constitutive relations to describe the rheological behavior.

In order to model the behavior at weakly nonlinear strains, the rate of strain tensor in Eq. (4) is replaced by a finite rate of strain measure $\partial \underset{\approx}{\mathbf{C}}^{-1}(t)/\partial t$, where $\underset{\approx}{\mathbf{C}}^{-1}(t)$ is the Finger strain tensor. This replacement results in a simplified frame invariant constitutive equation that belongs to the family of K-BKZ (Kaye-Bernstein-Kearsley-Zapas) model, taking the form [39]:

$$\underset{\approx}{\sigma}(t) = -\int_0^t G(t-t') \frac{\partial \underset{\approx}{\mathbf{C}}^{-1}(t')}{\partial t'} dt'. \tag{6}$$

In the limit of small strains, Eq. (6) leads to the linear viscoelastic constitutive Eq. (4). This class of K-BKZ equation has been used to model the nonlinear viscoelastic properties of different polymer gels [19,40,41].

In contrast to the linear response regime, as strain amplitude becomes nonlinear, the relaxation modulus is observed to be dependent on time as well as the magnitude of strain represented as: $G = G(t,\gamma)$. Interestingly, this dependence can be factorised into time and strain-dependent terms as: $G(t,\gamma) = G(t)h(\gamma)$, where $h(\gamma)$ is known as the damping function. As suggested by Wagner and Meissner [42], $h(\gamma)$ denotes the survival probability of the network on deformation. Mathematically, $h(\gamma)$ is bounded between $0 \leq h(\gamma) \leq 1$, where the lower limit signifies complete rupture of the network junctions at large strain amplitudes while the upper limit denotes the linear response regime. The damping function, therefore, characterizes the nonlinear response of the material. Incorporating the damping function into the integral Eq. (6) leads to Wagner equation [43], which is a special case of separable K-BKZ model, given by [44]:

$$\underset{\approx}{\sigma}(t) = -\int_0^t G(t-t')h(t,t',\gamma) \frac{\partial \underset{\approx}{\mathbf{C}}^{-1}(t')}{\partial t'} dt'. \tag{7}$$

Many analytical expressions for the damping function have been reported in the literature for different viscoelastic materials [45]. Wagner [43] first represented $h(\gamma)$ for low-density polyethylene (LDPE) melt as $h(I_2) = \exp(-k\sqrt{I_2 - 3})$, where $I_2 = \frac{1}{2}\left[\left(\operatorname{tr}\underset{\approx}{C}^{-1}\right)^2 - \operatorname{tr}\left(\underset{\approx}{C}^{-1}\right)^2\right]$ is the second invariant of the Finger strain tensor and $k$ is the material parameter. Wagner [43] observed that Eq. (7) correctly predicted the transient shear viscosity in LDPE melt for the low value of shear rates. Laun [40] modified the damping function by adopting a sum of two exponentials, which led to a better description of transient stress evolution in LDPE melt. Larson and Valesano [46] studied the stress response upon the application of double step strain to the LDPE melt by incorporating $h(\gamma)$ into the constitutive equation, however the prediction failed to capture the experimental behavior. In addition to shear flow, damping function also provides a theoretical framework in modelling the elongational flow in polymer melts [43,47]. However, it is important to note that these polymer melts did not exhibit a power-law distribution of relaxation time as observed for the critical gel state. Ng *et al.* [35] worked on critical gels of wheat dough during extension and shear flow. The incorporation of damping function given by: $h(\gamma) = 1/(1 + q\gamma^{2k})$, where $q$ and $k$ are the fitting parameters, predicted the transient uniaxial extensional data in wheat dough. Very recently, Keshavarz *et al.* [19] described the stress response of the polymer gel (casein network at the critical state) for a start-up shear flow, wherein $h(\gamma)$ exhibited both the strain hardening and softening behavior. While some work has been carried out to study nonlinear rheological behavior of the polymer gels [12,19] and food gels [35] at the critical state, no such work has been carried out on the colloidal gels at the critical point for which the interparticle interactions are much weaker than the chemical gel.

In this work, we study the linear viscoelastic behavior and transition to nonlinearity of a colloidal gel formed by a synthetic hectorite clay at the critical state. The objective of this work is threefold. Firstly, we demonstrate the validation of the Winter criteria for the critical gel state of the colloidal gel. Secondly, we report creep and recovery, stress relaxation, and start-up shear experiments on the colloidal gel at the critical state and prediction of the same using the constitutive Gel equation. Lastly, we investigate the weakly nonlinear behavior and develop a constitutive formalism to predict the same from a linear viscoelastic framework.

## II. Material and Experimental Procedure:

We employ a model synthetic hectorite clay LAPONITE® XLG (a registered trademark of BYK Additives). The primary particles of the same are disc-shaped having a diameter in the range of 25-30 nm and a thickness of about 1 nm [48]. The synthetic hectorite clay is a 2:1 layered silicate with an octahedral layer of magnesia sandwiched between two tetrahedral layers of silica [49]. The chemical formula of its unit crystal is given by: $\mathrm{Na}_{+0.7}\left[\left(\mathrm{Si}_8\mathrm{Mg}_{5.5}\mathrm{Li}_{0.3}\right)\mathrm{O}_{20}\left(\mathrm{OH}\right)_4\right]_{-0.7}$. Owing to the dissimilar charges on synthetic hectorite clay particle, the particles share attractive and repulsive interactions in the aqueous medium. As a result, the microstructure evolves continuously in order to achieve a progressively lower free energy state [48].

In this work, we prepare 3 weight % LAPONITE® XLG, 2 mM NaCl dispersion by mixing the oven dried (120 °C for four hours) clay in Millipore water (resistivity 18.2 MΩ cm) having predetermined quantity of salt (NaCl). Stirring is carried out by IKA-Ultra Turrax drive for 30 minutes leading to a transparent dispersion. In all the experiments, the freshly prepared samples are used to carry out the rheometric studies immediately after the stirring is over. The shear rheometry is performed using Dynamic Hybrid Rheometer 3 (TA Instruments). For each experiment, a fresh sample is loaded on a concentric cylinder geometry having a cup diameter of 30 mm and a gap of 1 mm. Subsequently, a cyclic frequency sweep experiment is conducted at a constant stress of 0.1 Pa over an angular frequency range of 0.5-25 rad/s. In order to ensure that the material properties do not evolve much during a single cycle, the frequency range is chosen in such a manner that the mutation numbers $\left(N'_{mu} = \left(2\pi/\omega G'\right)\left(\partial G'/\partial t\right)\right)$ and $\left(N''_{mu} = \left(2\pi/\omega G''\right)\left(\partial G''/\partial t\right)\right)$ remain within the specified limits: $N'_{mu} < 0.1$ and $N''_{mu} < N'_{mu}$ [50]. Furthermore, the colloidal gel at the critical state is probed by subjecting it to step strain (stress relaxation), step stress (creep) and step shear rate (start-up of shear flow) in the linear and nonlinear domain. In all the experiments a thin layer of silicon oil is applied to the free surface to prevent evaporation losses and the experiments are carried out at a constant temperature of 30 °C maintained by a Peltier system.

## III. Results and Discussions:

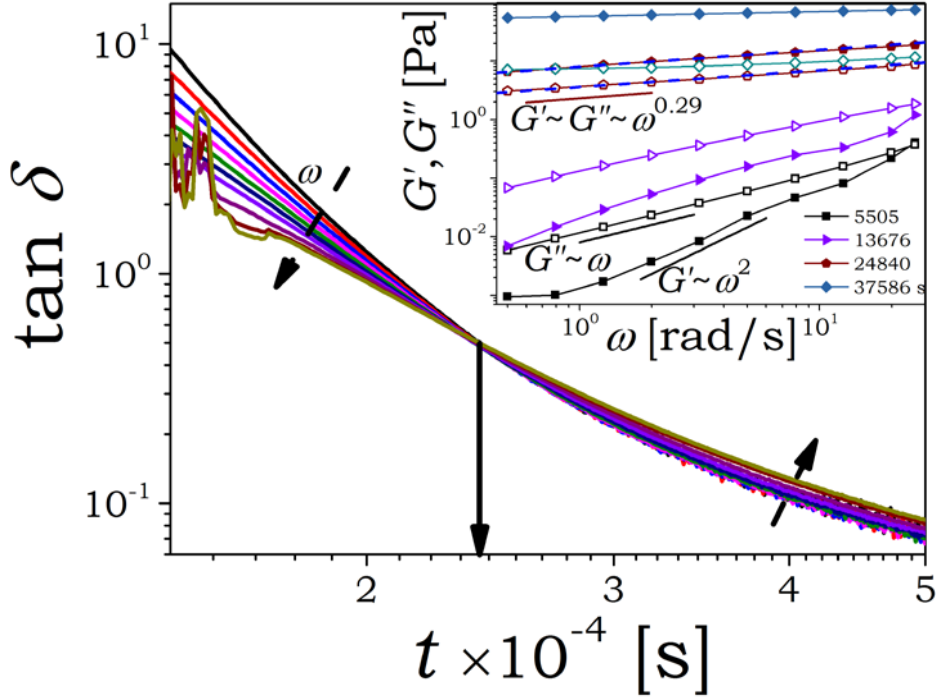

**FIG. 1.** Evolution of $\tan\delta$ is plotted as a function of time at different frequencies during the cyclic frequency sweep measurement. The point at which all the iso-frequency curves intersect represents the critical gel state. The dashed arrows indicate the direction of increasing values of $\omega$ such that in the sol (pre-gel) state $\tan\delta$ decreases with frequency, while in the post-gel state $\tan\delta$ increases with frequency. The inset shows $\omega$ dependence of $G'$ (closed symbols) and $G''$ (open symbols) at different times. At the critical gel state, both $G'$ and $G''$ show an identical power-law dependence on frequency given by: $G', G'' \sim \omega^{0.29}$. The lines in the inset serve as a guide to the eye.

Oscillatory shear experiments have been carried over two decades of angular frequency $(\omega)$ on spontaneously aging aqueous synthetic hectorite clay dispersion. A freshly prepared clay dispersion has been subjected to a succession of frequency sweep experiments. The corresponding time evolution of frequency dependence of $G'$ and $G''$ is plotted in the inset of Fig. 1. At the beginning of the experiment, we observe the dominance of $G''$ over $G'$ across much of the frequency range with $G' \sim \omega^2$ and $G'' \sim \omega$ dependence, which is a characteristic feature of the terminal regime suggestive of the sol (liquid) state. With an increase in time, both the moduli grow, however, the rate of growth of $G'$ is higher than $G''$. Eventually, the growth index decreases below 1 and $G'$ exceeds $G''$. At a certain time, both the moduli show identical dependence on frequency $\left(G' \sim G'' \sim \omega^{0.29}\right)$ as described by Eq. (1), which is the rheological signature of a critical gel state. In Fig. 1, we plot the evolution of $\tan \delta$ with respect to time $t$ at different frequencies. The scaling of the dynamic moduli in the inset suggests $\tan \delta$ to be inversely proportional to $\omega$ in the sol state as observed in the Fig. 1, wherein the dashed arow indicates the direction of increasing value of $\omega$. The critical gel state is identified by the intersection of all iso-frequency $\tan \delta$ curves at the identical point, thus making it independent of observation timescale [27]. The corresponding time is termed as the critical gelation time $(t_g)$, which for the present system is $t_g = 6.9$ hours as indicated by the solid arrow in Fig. 1. The relaxation exponent $(n_c)$ at the critical gel state, given by: $n_c = 2\delta/\pi$, is same as the power-law exponent of the moduli dependence on $\omega$, and is observed to be $n_c = 0.29$. As proposed by Muthukumar [36], the corresponding fractal dimension of the critical state is $f_d = 2.24$, which suggests the present system to undergo reaction limited aggregation. With the increase in time after the critical state, the dependence of dynamic moduli on $\omega$ continues to weaken such that $G'$ and $G''$ no longer remain parallel to each other. Furthermore, at higher times, $\tan \delta$ increases with an increase in $\omega$ that is suggestive of the post (critical) gel state. Thus Fig. 1 clearly corroborates the Winter criteria [3,27] for the critical gel transition in 3 weight% synthetic hectorite 2 mM NaCl dispersion.

The analysis of the linear viscoelastic behavior therefore suggests that an aqueous dispersion of synthetic hectorite clay shows all the characteristic features of the sol-gel transition observed for polymeric systems undergoing a crosslinking reaction. This work therefore suggests remarkable universality associated with the percolated state that gets validated not just for the chemical but also for the physical colloidal gels. However, from an application point of view, the materials are exposed to nonlinear deformations, which cannot be analyzed using linear viscoelasticity principles. Hence there is a need for a viscoelastic constitutive relation that can account for the nonlinear deformations.

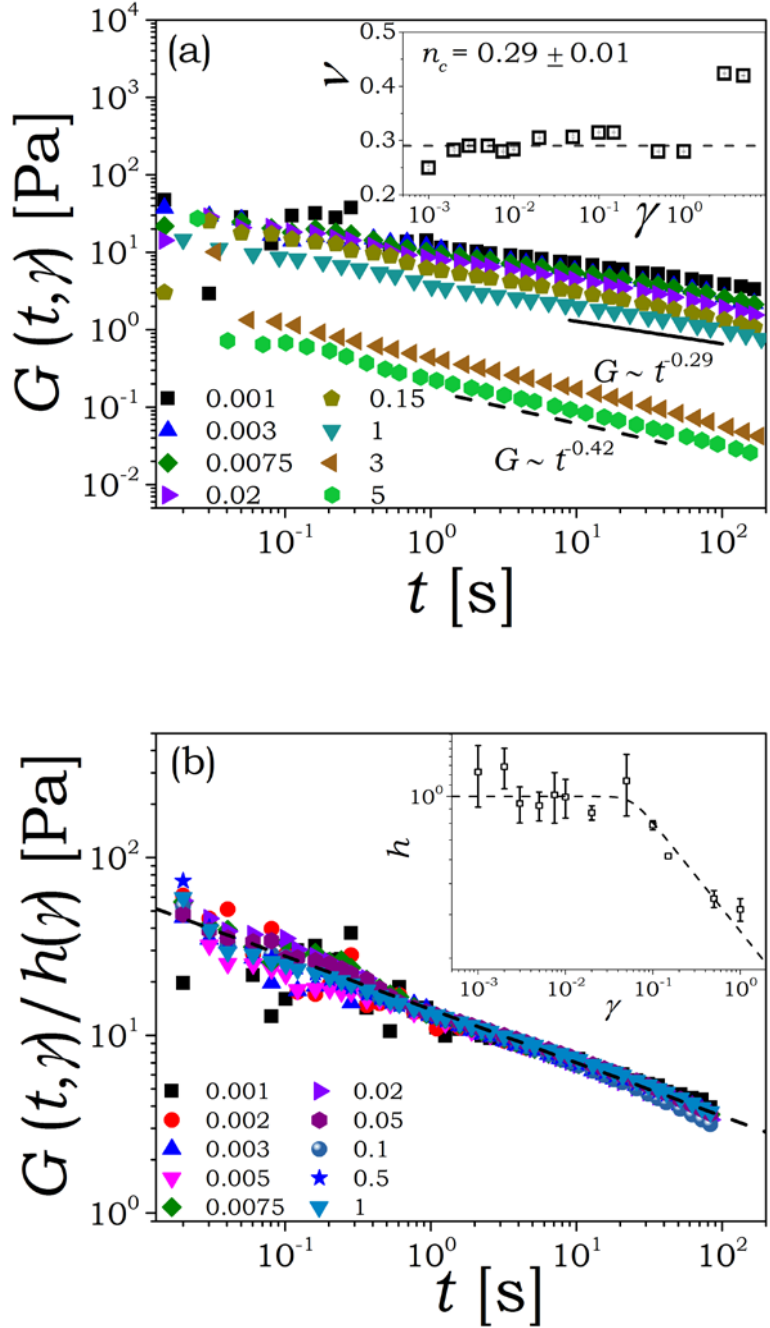

**FIG. 2.** (a) Evolution of relaxation modulus $G$ is plotted as a function of time. In all the experiments, the step strain (with magnitudes between 0.001 and 5) is applied at the critical point. The inset denotes the variation of exponent $\nu$ obtained from the power-law fit to $G$ data ($G \sim t^{-\nu}$). The dashed line indicates the mean value of critical exponent of 0.29 in the linear and weakly nonlinear region. (b) Construction of master curve obtained by vertically shifting $G(t,\gamma)$ data at each strain by a damping function $h(\gamma)$. The solid line represents a power-law fit to the superposed data with a slope of – 0.29. The inset shows the dependence of $h(\gamma)$ on strain as obtained by vertically shifting data shown in (a). The dashed line denotes the functional fit of Eq. (10) to $h(\gamma)$ data.

We now investigate the linear and nonlinear behavior of the present system at the critical gel state. In order to ensure the presence of a critical gel state, a cyclic frequency sweep experiment is performed. We stop the cyclic frequency sweep experiment at the point where $\tan\delta$ becomes independent of frequency. Since each experiment requires a fresh sample, the extent of achieving the critical gel state varies slightly from sample to sample within an experimental uncertainty. We subject the critical gel state to various deformation fields over a short period of time ($t << t_g$), therefore assuming that over the duration of application of deformation field, the system remains at the critical state.

We first subject the critical state to the step strain experiments at different magnitudes of strain $(\gamma)$ over a range of 0.001 to 5. In Fig. 2(a), we plot time evolution of relaxation modulus ($G(t,\gamma)$) at different magnitudes of strain. For clarity, only few data sets corresponding to step strain have been shown in Fig. 2(a). It can be seen that irrespective of the magnitude of the strain, the relaxation modulus shows a power-law dependence on time $G(t,\gamma) \sim t^{-\nu}$. In the inset of Fig. 2(a) we plot the power-law coefficient $\nu$ as a function of $\gamma$ for all the values of step strain. On one hand, the relaxation modulus data get superposed on each other, within an experimental uncertainty, only up to small values of strain corresponding to the linear response regime. On the other hand, $\nu$ remains very close to that of the critical value ($\nu = n_c = 0.29 \pm 0.01$) over a larger range of strains that is up to $\gamma \approx 1$. Over the range of strains, for which although $\nu$ shows the same value, magnitude of $G(t,\gamma)$ decreases with an increase in $\gamma$. We represent this regime as weakly nonlinear. For higher strains ($\gamma > 1$), there is a steep increase in the power-law coefficient to about $\nu \approx 0.42$ indicated by dashed line in Fig. 2(a), suggesting this regime to be strongly nonlinear.

The independence of decay of relaxation modulus on strain in the linear response regime and its decrease at higher strains while preserving the curvature is observed for many viscoelastic systems [12,19,26]. The self-similar curvature of $G(t,\gamma)$ facilitates vertical shifting by using a damping function $h(\gamma)$ leading to a superposition as shown in Fig. 2(b). Since the very fact that superposition exists, we can factorize $G(t,\gamma)$ into a time-dependent $G(t)$ and strain-dependent term $h(\gamma)$ given by [12]:

$$G(t,\gamma) = G(t)h(\gamma). \tag{8}$$

For a critical gel, the Eq. (8) reduces to:

$$G(t,\gamma) = St^{-n_c}h(\gamma), \tag{9}$$

for $\gamma < 1$. In the inset of Fig. 2(b) we plot the damping function as a function of $\gamma$, which suggests that the linear viscoelastic region is up to $\gamma_l \approx 0.059$, wherein the absence of strain dependence leads to $h = 1$. With the increase in $\gamma$ from $\gamma_l$ to the large strains (up to $\gamma \approx 1$), $h(\gamma)$ decreases suggesting softening of the critical gel with an increase in $\gamma$. Considering the nature of the damping function, we fit Carreau Yasuda type expression given by:

$$h(\gamma) = \left[1 + \left(\frac{\gamma}{\gamma_l}\right)^a\right]^{-m/a}, \tag{10}$$

where $\gamma_l = 0.059$ is the magnitude of strain associated with the linear response regime while $m = 0.44$ and $a = 5.243$ are the model parameters. It can be seen that Eq. (10), represented by the dotted line in the inset of Fig. 2(b), fits the experimental data very well.

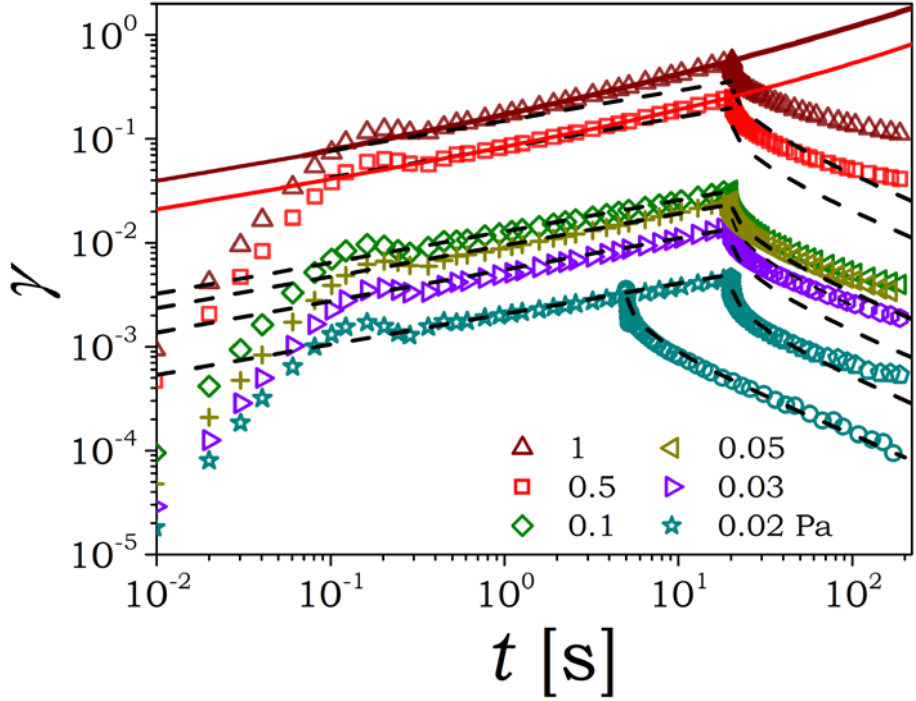

**FIG. 3.** Evolution of shear strain $(\gamma)$ is plotted as a function of time subsequent to application of constant shear stress at the critical gel state and during recovery after withdrawal of the shear stress. The dotted lines are the predictions of shear strain from the Gel equation (13) for a linear deformation. The solid line is predicted using the modified expression (Eq. (14)) of the constitutive equation for a weakly non–linear deformation.

Next we study the stress controlled response wherein we subject the critical gel to a constant value of stress followed by the recovery. The corresponding strain response is plotted in Fig. 3 over a broad range of stresses (0.02 to 1 Pa). On application of stress, at short times ($t < 0.5$ s), the measured value of strain undergoes damped oscillation resulting from coupling between the sample elasticity and the instrument inertia [15,51]. The values of strain upon attenuation of oscillations suggest instantaneous deformation in the material that follows a gradual increase. After a certain creep time, the sample is allowed to recover by putting stress equal to zero. The sample shows an instantaneous strain recovery followed by a delayed recovery; however the induced strain does not recover completely as shown in Fig. 3. The applied stress-controlled deformation field can be represented as:

$$\sigma(t) = \sigma_0 [u(t) - u(t-t_c)], \tag{11}$$

where $\sigma_0$ is the constant creep stress, $t$ is the experimental timescale, $t_c$ is the creep time, and $u(t)$ is the unit step function. The strain induced in the material can be obtained by using the Boltzmann superposition principle (BSP) given by [28]:

$$\gamma(t) = \int_0^t J(t-t') \frac{d\sigma}{dt'} dt'. \tag{12}$$

Incorporation of Eq. (3) for $J(t)$ into Eq. (12) and solving it for the deformation field given by Eq. (11) leads to:

$$\gamma(t) = \frac{(1-n_c)\sigma_0}{S\Gamma(2-n_c)\Gamma(1+n_c)} \left( t^{n_c} u(t) - (t-t_c)^{n_c} u(t-t_c) \right). \tag{13}$$

In Fig. 3, Eq. (13) is plotted as the dashed lines for the various creep stresses. It can be seen that for the creep stresses below 0.1 Pa, Eq. (13) shows an excellent prediction of the experimental data. Equation (13) also shows a good prediction of the recovery data when the total creep and recovery time is sufficiently small (refer to data associated with $\sigma_0 = 0.02$ Pa and creep time of 5 s). However, with an increase in recovery time, Eq. (13) overpredicts the recovery compared to the experimental data. We feel that this deviation is due to continuous gelation of the present system with increase in time, which takes the system into the post-gel state with progressively greater crosslinking density causing recovery to be lower than that predicted for the critical gel state.

As shown in Fig. 3, upon application of the higher values of stress (0.5 Pa and 1 Pa), the strain induced in the material at the critical state deviates from the prediction of Eq. (13) suggesting material to undergo nonlinear deformation. The observed deviation in strain response could be due to the beginning of rupture of the percolated network associated with the critical state. This behavior is qualitatively similar to stress relaxation associated with the higher magnitudes of step strain as expressed by the values of damping function below unity. Incorporating $h(\gamma)$ given by Eq. (10) and the relaxation modulus given by Eq. (2) into Eq. (7) for a creep flow with constant stress $\sigma_0$, we get:

$$\sigma_0 = S\int_0^t (t-t')^{-n_c}\left[1+\left(\gamma/\gamma_l\right)^a\right]^{-m/a}\frac{d\gamma}{dt'}dt'. \tag{14}$$

We propose that, on application of higher stress for a longer duration, the extent of breakage of the network continuously increases. As a result, the rate of growth in strain constantly increases with time. In order to accommodate this effect, we solve Eq. (14) for a creep flow under application of constant stress $\sigma_0$ to obtain strain as a function of time. Since Eq. (14) is implicit in strain as well as strain rate, that also inside the integral, it can be solved only by assuming a suitable form of strain. We propose a form similar to that given by Eq. (3) but with a strain dependent pre-factor and relaxation exponent represented by:

$$\gamma = \overline{A}(\gamma)t'^{n(\gamma)} = \overline{A}(\gamma)t^{n(\gamma)}x^{n(\gamma)}. \tag{15}$$

where $\overline{A}(\gamma) = A = \dfrac{\sigma_0(1-n_c)}{S\Gamma(2-n_c)\Gamma(1+n_c)}$ and $n(\gamma) = n_c$ for $\gamma \leq \gamma_l$. In Eq. (15) we normalize time ($t'$) by the present time $t$, which is a constant in the integral given by Eq. (14) leading to $x = t'/t$. This is because, owing to the lack of characteristic relaxation time associated with the critical state, the present time $t$ becomes a natural choice for nondimensionalization.

We represent the strain dependent relaxation exponent $n(\gamma)$ using the Taylor series expansion centered at $\gamma_l$ as:

$$n(\gamma) = n_c + \left(\frac{dn(\gamma)}{d\gamma}\bigg|_{\gamma=\gamma_l}\right)(\gamma-\gamma_l) + O(\gamma^2). \qquad \ldots \gamma \geq \gamma_l \tag{16}$$

Similarly, the prefactor $\overline{A}(\gamma)t^{n(\gamma)}$ can be expanded using Taylor series as:

$$\overline{A}(\gamma)t^{n(\gamma)} = At^{n_c} + \left(\frac{d\overline{A}(\gamma)t^{n(\gamma)}}{d\gamma}\bigg|_{\gamma=\gamma_l}\right)(\gamma-\gamma_l) + O(\gamma^2). \quad \ldots \gamma \geq \gamma_l \quad (17)$$

In the above expression, we expand $\overline{A}(\gamma)t^{n(\gamma)}$ and not $\overline{A}(\gamma)$ using the Taylor series because dimensions of $\overline{A}(\gamma)$ depend on $\gamma$ in such a way that $\overline{A}(\gamma)t^{n(\gamma)}$ is dimensionless. In Eqs. (16) and (17), in principle, the strain should have the actual time dependence given by Eq. (14). However since Eq. (17) requires the difference: $(\gamma-\gamma_l)$, we approximate $\gamma$ to be given by the linear expression (Eq. (3)) and ignore the higher order terms to facilitate numerical solution without much loss of accuracy. By incorporating Eq. (13) (only for creep flow) into Eqs. (16) and (17) and nondimensionalizing we get:

$$n[\gamma(x)] = n_c + \alpha\left(\frac{x^{n_c}}{x_0^{n_c}} - 1\right) \text{ and} \quad \ldots x \geq x_0 \quad (18)$$

$$\overline{A}[\gamma(x)]t^{n[\gamma(x)]} = At^{n_c} + C\left(\frac{x^{n_c}}{x_0^{n_c}} - 1\right), \quad \ldots x \geq x_0 \quad (19)$$

where $x_0$ represents the normalized time associated with $\gamma_l$ leading to $x_0 = t_0/t$ and the pre-factors are simplified as $\alpha = \left(\frac{dn(\gamma)}{d\gamma}\bigg|_{\gamma=\gamma_l}\right)\gamma_l$ and $C = \left(\frac{d\overline{A}(\gamma)t^{n(\gamma)}}{d\gamma}\bigg|_{\gamma=\gamma_l}\right)\gamma_l$.

On substitution of Eqs. (15), (18) and (19) in Eq. (14), we get the relationship between the constant creep stress and the strain for a nonlinear deformation. We solve Eq. (14) numerically to get the model parameters $\alpha$ and $C$ for different values of creep time $t$ and $\sigma_0$. For a creep time of $t = 20$ s and $\sigma_0 = 1$ Pa, we get $\alpha = 0.0045$ and $C = 0.05$ while for $t = 20$ s and $\sigma_0 = 0.5$ Pa, the value comes out to be $\alpha = 0.017$ and $C = 0.033$. It should be noted that the values of $\alpha$ and $C$ are solely obtained from the value of $\sigma_0$. The pre-factor $\alpha$, that determines the curvature of the strain evolution in the nonlinear domain, has been observed to be independent of $t$. However, since $C$ is associated with the first derivative of $\overline{A}(\gamma)t^{n(\gamma)}$ with respect to $\gamma$, it is observed to increase with $t$. The predictions of Eq. (14) are plotted as the continuous lines in Fig. 3. It can be seen that the prediction matches the experimental data very well. It is important to note that even though the value of $C$ increases with $t$, the strain evolution curves identically coincide with each other for different values of $t$, as it should be, in order to have internal consistency. This clearly suggests that the KBKZ model and the semi-empirical expression given by Eqs. (15), (18) and (19) capture the evolution of strain in the nonlinear region very well.

The extent of breakage of the network that the material undergoes during the nonlinear creep can be expressed in terms of energy required for dissociation of the percolated network. The energy per unit volume required to induce strain $\gamma$ in a material during a creep experiment at stress $\sigma_0$ is: $e = \sigma_0 \gamma$ [24]. Assuming that the percolated network at the critical gel state does not break during the deformation in the linear viscoelastic limit; the energy required to induce nonlinearity at creep time $t$, which could be associated with the rupture of a network, is given by: $e(t) = \sigma_0 \left( \gamma^{\text{nonlinear}}(t) - \gamma^{\text{linear}}(t) \right)$. Therefore, as shown in Fig. 3, in a typical creep experiment at $\sigma_0 = 1$ Pa, the additional energy for the network breakage at the time when nonlinear strain reaches a maximum value is given by:

$$e = \sigma_0 \left( \gamma^{\text{nonlinear}}_{\max} - \gamma^{\text{linear}}_{\max} \right) = 0.23 \ \text{J/m}^3. \tag{20}$$

For polydimethylsiloxane (PDMS) gel near its critical point, the energy required for network rupture was calculated to be $3.64 \times 10^3 \ \text{J/m}^3$ [24], which is much higher compared to the synthetic clay dispersion. The PDMS gel is a stronger gel, where the breakage of the network occurs at 800 Pa, owing to the covalent bonds associated with the same. However, synthetic clay dispersion studied in this work deviates from linearity at very small stress of 1 Pa. This leads to the low value of $e$ for the present clay dispersion owing to weaker interparticle interactions. For the 3 weight % clay with 2 mM salt, the activation energy associated with the gelation process is given by $9800 \ \text{J/m}^3$ [7]. Therefore, the present discussion suggests an extremely small amount of mechanical energy is required to initiate breakage in the critical gel compared to the activation energy necessary to form it.

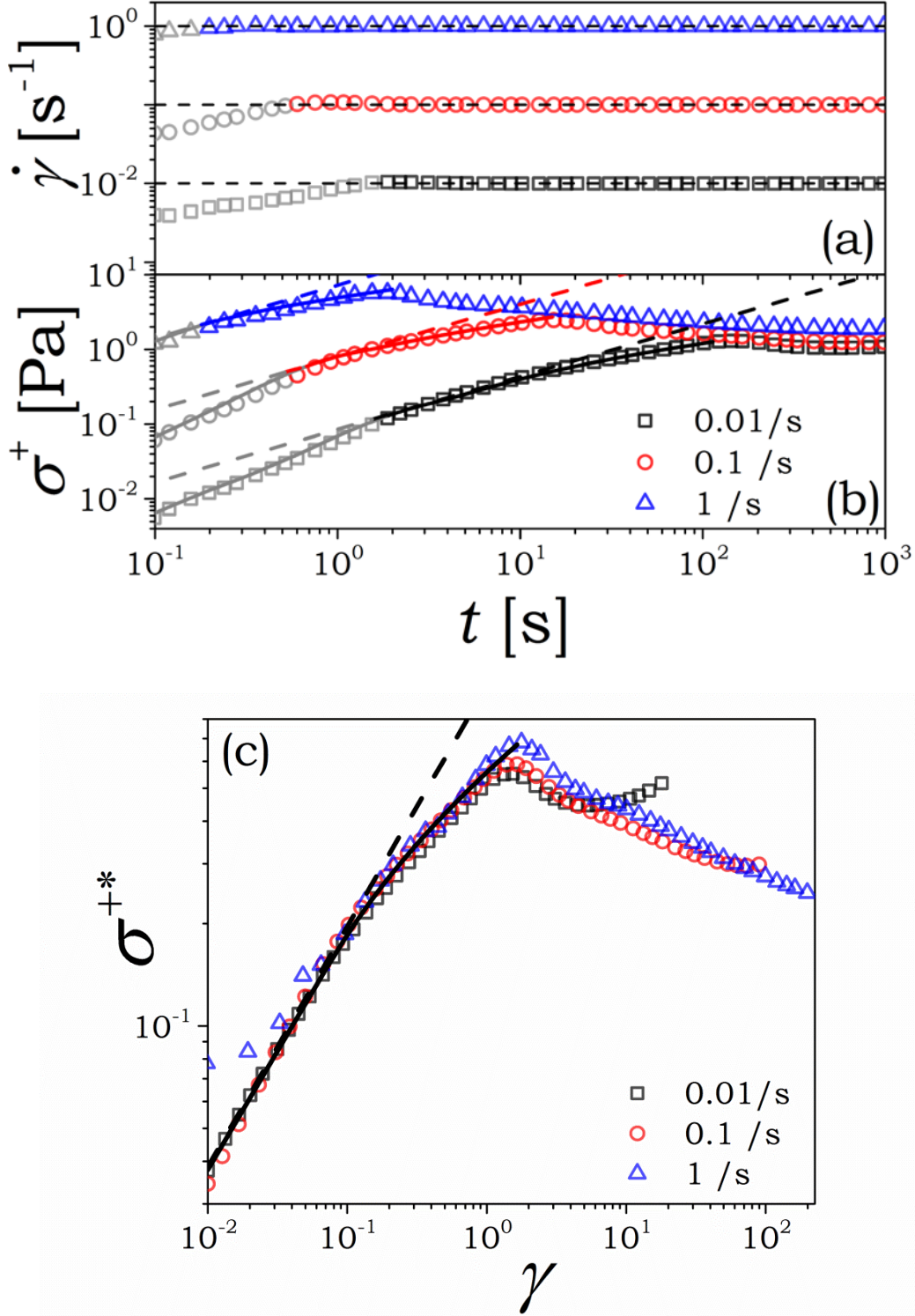

**FIG. 4.** (a) The growth of shear rate $(\dot{\gamma})$ is plotted as a function of time on the application of a step shear rate $(\dot{\gamma}_0)$ during the start-up of shear flow experiment.

The dashed line denotes the constant value of $\dot{\gamma}_0$ provided as an input to the rheometer. (b) The evolution of transient shear stress $\left(\sigma^+\right)$ is plotted as a function of time $(t)$ for different step shear rates $\left(\dot{\gamma}_0\right)$ applied at the critical gel point. The dashed line is the prediction of $\sigma^+$ in a limit of linear viscoelasticity using Eq. (21) while the solid line is the prediction using the modified expression given by Eq. (23). (c) The normalized transient stress $\left(\sigma^{+*}\right)$ is plotted as a function of shear strain $(\gamma)$ for the same data shown in (b). The dashed line denotes the linear response given by Eq. (22) and the solid line is the prediction of $\sigma^{+*}$ from Eq. (23).

We now study the response of the material at the critical gel state upon an application of step shear rate $\left(\dot{\gamma}_0\right)$. The evolution of $\dot{\gamma}$ on application of a step shear rate is plotted as a function of time during the start-up shear flow. It can be seen that $\dot{\gamma}$ undergoes oscillation before achieving a constant value of $\dot{\gamma}_0$. The time required by the rheometer to achieve a constant pre-set value of $\dot{\gamma}_0$ increases as we decrease the magnitude of step shear rate. Hence, we are limited by the instrument in exploring shear rates below 0.01/s for the critical gel studied in the present work. In Fig. 4(b), the evolution of transient stress $(\sigma^+)$ is plotted as a function of time during a start-up shear experiment at the critical gel state for three different but constant values of shear rates ($\dot{\gamma}_0 =$ 0.01, 0.1 and 1/s). It can be seen that $\sigma^+$ increases with time in a limit of small times. At certain time, the evolution of $\sigma^+$ undergoes a deviation from the linear behavior, and eventually it reaches a maximum value. Subsequently, it decreases with increase in time. As expected, the stress evolution gets shifted to higher values with increase in $\dot{\gamma}_0$. Upon application of the step strain rate $\left(\dot{\gamma}_0\right)$ at the critical gel state, the stress $\left(\sigma^+\right)$ induced in the material can be obtained by using Eq. (4) and is given by:

$$\sigma^+(t) = \int_0^t S(t-t')^{-n_c} \frac{d\gamma}{dt'} dt' = \frac{S\dot{\gamma}_0}{1-n_c} t^{1-n_c}. \tag{21}$$

The theoretical prediction of $\sigma^+$ given by Eq. (21), which is shown as dashed line in Fig. 4(b), matches the experimental data very well. Since the shear rate is constant, the strain is given by $\gamma = \dot{\gamma}_0 t$. Consequently in Eq. (21), $\dot{\gamma}_0 t^{1-n_c}$ can be replaced by $\dot{\gamma}_0^{n_c} \gamma^{1-n_c}$. Upon further adjustment, Eq. (21) leads to:

$$\sigma^{+*}(\gamma) = \frac{1-n_c}{S\dot{\gamma}_0^{n_c}} \sigma^+(\gamma) = \gamma^{1-n_c}. \tag{22}$$

In Fig. 4(c), we represent the evolution of normalized stress $(\sigma^{+*})$ as a function of strain $(\gamma = \dot{\gamma}_0 t)$ using the dashed lines for the data shown in Fig. 4(b). Interestingly, all the $\sigma^{+*}$ evolution data superpose not just in the linear domain but also until $\sigma^{+*}$ reaches a maximum value irrespective of the shear rate.

In Fig. 4(c), it can be seen that the prediction matches the experimental data very well in the linear viscoelastic region ($\gamma < 0.059$). With an increase in the value of strain beyond linear domain ($\gamma > 0.059$), however, the experimental data deviates from the linear dependence and is observed to be lower in magnitude than the linear prediction. It is interesting to note that such divergence from the linear behavior was not observed in similar start-up shear experiment on controlled PDMS samples close to the gel point [24]. At a higher value of strain, beyond the maximum in stress, the material undergoes macroscopic failure of the critical gel network. At this point, the structure breaks down completely and the material starts flowing. Therefore, similar to that observed in Fig. 2(a), the complete behavior can be divided into three regimes: linear $(\gamma < 0.059)$, weakly nonlinear $(0.059 < \gamma < 1)$ and nonlinear $(\gamma > 1)$. The dynamics of the linear regime is well explained by Eq. (22). On the other hand, the rheological behavior of the material in the nonlinear regime is itself a complex problem, and is beyond the scope of the present work. In order to model the deviation from linearity, we incorporate relaxation modulus given by Eq. (2) into Eq. (7) for a constant shear rate to give:

$$\sigma^+(t) = S\dot{\gamma}_0 \int_0^t (t-t')^{-n_c} \left[1 + \left(\dot{\gamma}_0(t-t')/\gamma_l\right)^a\right]^{-m/a} dt'. \tag{23}$$

We solve Eq. (23) numerically and the corresponding prediction of $\sigma^{+*}$ without any adjustable parameter is shown by a solid line in Fig. 4(c). Very interestingly, the quasi-linear model shows a remarkable prediction of the experimental data of the fractal critical gel not just in the linear regime but also in the weakly nonlinear regime with just a knowledge of damping function. However, the proposed model does not account for the decrease in $\sigma^{+*}$ after the macroscopic fracture of the network since the model is based on the assumption of uniform strain field [19].

In the literature, the rupture dynamics of the percolated network has been described using the Bailey durability criteria [19,52], which represents the mechanical failure in brittle solids and glasses [52]. If $t_f$ denotes the time to rupture of the network in a certain deformation process, the Bailey criterion of durability is expressed as [52]:

$$\int_0^{t_f} \frac{dt}{\tau(\sigma(t))} = 1, \qquad (24)$$

where $\tau$ is the durability function that depends on the magnitude of stress a material is subjected to. The durability function signifies the dependence of the rupture time on the deformation field. In the literature, the durability function is proposed to have a power-law dependence on the constant stress in the creep experiments on polymeric critical gels given by [19]: $\tau(\sigma_0) = B\sigma_0^{-\beta}$. Incorporation of this dependence in Eq. (24) therefore leads to:

$$t_f = B\sigma_0^{-\beta}. \qquad (25)$$

If we assume that the deviation from linearity marks the initiation of the rupture of a network, the time to rupture can be computed using the point of departure from linearity (The point of rupture marks the end of linearity, and therefore it is the last point that obeys the linear behavior). The departure from linearity can be obtained in terms of stress from the start-up shear experiments. On substituting the time to rupture given by Eq. (25) into Eq. (21) and rearranging the equation, we get:

$$\sigma_l = \left(\frac{B^{1-n_c}S}{1-n_c}\right)^{\frac{1}{1+\beta(1-n_c)}} \dot{\gamma}_0^{\frac{1}{1+\beta(1-n_c)}}. \qquad (26)$$

The point of departure from linearity, which we consider as the point of initiation of rupture, can be independently obtained from linear viscoelastic principles. Since $\gamma_l$ marks the largest strain associated with the linear regime, incorporating $t_f = \gamma_l/\dot{\gamma}_0$ in Eq. (21) for a start-up of shear flow, the shear rate dependence of stress associated with deviation from linearity $\left(\sigma_l\right)$ can be expressed as:

$$\sigma_l = \frac{S\gamma_l^{1-n_c}}{1-n_c}\dot{\gamma}_0^{n_c}. \tag{27}$$

On comparing the scaling of $\sigma_l$ on $\dot{\gamma}_0$ given in Eqs. (26) and (27), we get the following relation:

$$\beta = \frac{1}{n_c}. \tag{28}$$

The dependence between the critical relaxation exponent $n_c$ and $\beta$, therefore, suggests a direct relationship between the rupture dynamics described by the Bailey criterion and the onset of nonlinearity given by linear viscoelasticity for the critical colloidal gel. The very fact that the experimental creep and start-up shear data respectively shown in figures 3 and 4 follows the principles of linear viscoelasticity exceedingly well, the present system indeed follows Eq. (28) and consequently the Bailey criterion when point of initiation of rupture is considered as departure from linearity. Interestingly, Keshavarz *et al.* [19] also validate relation given by Eq. (28) for the 4 weight % Casein gel at the critical state.

This work, therefore, clearly indicates that the quasi-linear model remarkably describes the rheological behavior in the limit of linear viscoelasticity and, very importantly, in the weakly nonlinear region. It is important to note that the analysis of the stress relaxation behavior leading to damping function facilitates the complete rheological description of the material behavior before the macroscopic breakage of the network structure. We believe that the quasi-linear model and the methodology described here can be extended to study the linear and weakly nonlinear viscoelastic behavior of the wide range of soft materials at the critical gel state.

### IV. Conclusion:

In this paper, we investigate the linear and weakly nonlinear viscoelastic behavior of a colloidal gel composed of the synthetic hectorite clay (LAPONITE®) dispersion at the critical point. In the oscillatory shear experiments, the temporal evolution of $G'$ and $G''$ demonstrates all the characteristic features of the sol-gel transition passing through a critical gel state that is suggestive of the weakest percolated fractal network structure. We also perform a comprehensive set of stress relaxation, creep-recovery and start-up of shear experiments at the critical gel state. The stress relaxation experiments conducted over a wide range of linear and nonlinear strain values lead to a superposed master curve when relaxation modulus is vertically shifted using a strain dependent damping function of the Carreau Yasuda form. The resulting time–strain separable relaxation modulus when incorporated in K–BKZ integral equation produces a quasi-linear constitutive equation. We observe that the K–BKZ model with the damping function predicts the shear start-up and the creep behavior of the critical gel in the linear and weakly nonlinear flow region very well. We also analyze energy and stress associated with the network breakage during the nonlinear deformation using the developed viscoelastic framework and the Bailey criterion.

**Acknowledgement:** We acknowledge financial support from the Science and Engineering Research Board (SERB), Department of Science and Technology, Government of India. We also thank Professor Norman Wagner for discussion and insightful suggestions.